\documentclass[twocolumn,prl]{revtex4}
\usepackage {graphicx}
\begin{document}
\title{Periodic Droplet Formation in Chemically Patterned Microchannels}
\author {Olga Kuksenok$^{1}$, David Jasnow$^{2}$, Julia Yeomans$^{3}$, Anna C. Balazs$^{1}$}
\affiliation{$^{1}$ Department of Chemical Engineering, University of
Pittsburgh, Pittsburgh PA 15261}
\affiliation{$^{2}$ Department of Physics and Astronomy, University of
Pittsburgh, Pittsburgh PA 15261}
\affiliation{$^{3}$ Theoretical Physics, University of Oxford, Keble Road, Oxford, OX1 3NP, U.K.}
\begin{abstract}
Simulations show that when a phase-separated binary AB fluid is driven to flow 
past chemically patterned substrates in a microchannel, the fluid exhibits 
unique morphological instabilities. For the pattern studied, these 
instabilities give rise to the simultaneous, periodic formation of 
monodisperse droplets of A-in-B and B-in-A. The system bifurcates between 
time-independent behavior and different types of regular, non-decaying 
oscillations in the structural characteristics. The surprisingly complex 
behavior is observed even in the absence of hydrodynamic interactions and 
arises from the interplay between the fluid flow and patterned substrate, 
which introduces non-linearity into the dynamical system.
\end{abstract}
\maketitle

Hydrophilic-hydrophobic patterning is used by a variety of biosystems to direct the motion 
of fluids at surfaces. For example, hydrophobically-hydrophilically patterned 
backs help desert beetles to capture water, and hydrophobic patches control 
water permeation in leaves \cite{Ralf}. This motif is also used to steer the 
motion and reaction of fluid droplets in liquid microchips \cite{1} and is 
being utilized to design self-cleaning substrates \cite{Ralf}. Despite the utility 
of these designs, there have been surprisingly few theoretical studies into the 
dynamics of fluid flow over chemically patterned surfaces. In this study, we examine 
a conceptually simple system where two partially miscible fluids, A and B, 
are mechanically driven (by an imposed shear) to flow past patterned surfaces within 
a microchannel (see Fig. 1). The system exhibits two distinct steady-states; however, 
in the transition between the two states, we uncover intricately complicated behavior, 
where monodisperse droplets of both A-in-B and B-in-A are formed periodically in time 
(as shown in Fig. 2), and the confined liquid displays regular, non-decaying oscillations 
in its structural characteristics. Furthermore, we isolate points where this system 
bifurcates between time-independent behavior and different types of oscillatory patterns. 
What is striking is that the observed phenomena occur even in the absence of hydrodynamic 
interaction; this is distinct from well-known instabilities in fluids \cite{2},\cite{3}.
Given that the system is relatively simple, the results suggest that complex transitions 
between well-defined steady-states may well be evident in a broad variety of dynamical systems.

\begin{figure}[htbp]
\centerline{\includegraphics[width=3.34in,height=1.78in]{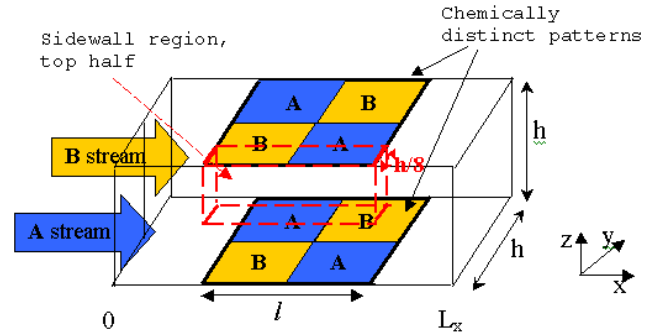}}
\label{fig1}
\caption{Schematic of system}
\end{figure}

The observed complex oscillatory patterns arise from a competition between 
advection and thermodynamics as an imposed Poiseuille flow drives the 
phase-separated fluids to flow over the chemically patterned substrates. As 
shown in Fig. 1, the top and bottom of the microchannel are decorated with a 
checkerboard pattern. Each checkerboard is composed of two A(B)-like 
patches, which are preferentially wetted by the A(B) fluid. The first B 
patch (in yellow) is placed in the way of the A stream (in blue) and 
correspondingly, the first A patch is located in the path of the B fluid.

The binary fluid is characterized by the order parameter 
$\varphi (r,t) = \rho _A (r,t) - \rho _B (r,t)$, where $\rho _i (r,t)$ 
represents the local number density of the \textit{i-th} component, $i = A,B$. The 
thermodynamic behavior of the system is governed by the coarse-grained free 
energy functional, $F = F_0 + \Psi _S $, where $F_0$ is the Ginzburg-Landau 
free energy for a binary mixture

\begin{equation}
\label{eq1}
F_0 = \int {d\vec {r}\left[ { - \frac{a}{2}\varphi ^2 + \frac{b}{4}\varphi 
^4 + \frac{k}{2}\left| {\vec {\nabla }\varphi } \right|^2} \right]} 
\end{equation}
\noindent
and $a$ and $b$ are positive constants. We consider the fluid to be in the 
two-phase coexistence regime where the equilibrium order parameter for the 
A(B) phase is $\varphi _{A / B} = \pm \sqrt {a / b} $. 
The term $\frac{k}{2}\left| {\vec {\nabla }\varphi } 
\right|^2$ represents the cost of order parameter gradients. The free energy 
$\Psi _s $ describes the interaction of a fluid element at a point $\vec 
{r}$ with the patterned substrate. Specifically, we take \cite{6}

\begin{equation}
\label{eq2}
\Psi _s = \int {d\vec {r}\int {d\vec {s}\left( {\frac{1}{2}V(\vec {s}) \cdot 
e^{ - \left| 
{\mathord{\buildrel{\lower3pt\hbox{$\scriptscriptstyle\rightharpoonup$}}\over 
{r}} - 
\mathord{\buildrel{\lower3pt\hbox{$\scriptscriptstyle\rightharpoonup$}}\over 
{s}} } \right| / r_0 }\left( {\varphi (\vec {r}) - \tilde {\varphi }(\vec 
{s})} \right)^2} \right)} } \quad,
\end{equation}
\noindent
where the inner integral represents integration over the substrates. $V(\vec 
{s}) = V = const$ on the patterns and is zero otherwise \cite{7}, and $r_0 $ 
represents the range of the substrate potential. We choose $\tilde {\varphi 
}(\vec {s}) = \varphi _{A(B)} $ to introduce $A(B)$ -wetted patches at specific regions of the substrate. 
Through eq (\ref{eq2}), the free energy $F$ is reduced when the fluid is A(B)-rich near A(B)-like patches.
\begin{figure}[htbp]
\centerline{\includegraphics[width=3.43in,height=2.15in]{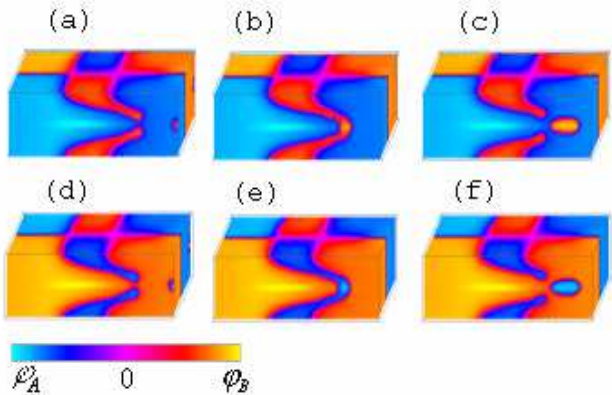}}
\label{fig2}
\caption{Periodic droplet formation. Blue represents A-rich and 
yellow represents B-rich domains. Panels a-c show the "front" and panels 
d-f show the "back" of the channel, with droplets of \textit{B-in-A} and \textit{A-in-B}, 
respectively. a, d) $t = 31200$, b, e) $t = 32000$, c,f) $t = 33600$. The other parameters 
are: $H = 3 \cdot 10^{ - 4}$, $l = 60$, $h = 40$, $r_0 = 5$, $V = 0.003$}
\end{figure}
The evolution of the order parameter for this system is described by the 
Cahn-Hillard equation, in which the flux of $\varphi $ is proportional to 
the gradient of the chemical potential, $J_{\varphi} = M \vec\nabla\mu$, 
where $\mu= \frac{\delta F}{\delta\varphi}$ and M is the mobility of the order parameter. 
The imposed Poiseuille flow advects variations in $\varphi 
$ along the microchannel. In dimensionless units \cite{8}

\begin{equation}
\label{eq3}
\frac{\partial \varphi }{\partial t} + \vec {v} \cdot \vec {\nabla }\varphi 
= \nabla ^2\mu 
\end{equation}
where the length scale is chosen to be equal to the thickness of the interface between A and B 
fluids, $\xi _{int} = \sqrt {k / a} $, and the scale of time is the diffusion time through that interface, 
$\tau = {\xi_{int}}^2 /aM$ \cite{8}.
The velocity field $\vec {v}$ obeys the Navier-Stokes equation in 
the over-damped limit, which is appropriate for low Reynolds number flow,
\begin{equation}
\label{eqNS}
0 = - \vec {\nabla }p + \nabla ^2\vec {v} + \vec {H}
+ C\frac{\delta F}{\delta \varphi}\vec {\nabla }\varphi 
\end{equation}
where p is a Lagrange multiplier that guarantees the incompressibility condition, 
$\vec {\nabla } \cdot \vec {v} = 0$, and $H_x = (P_{in}- P_{out})\xi _{int} \tau \eta /L $ is 
the dimensionless form of the imposed pressure drop $(P_{in}- P_{out})$ along the channel of length $L$.
Because the pressure gradient is applied along the x-axis, only the x-component of the vector $\vec {H}$ 
is non-zero, $H_x \equiv H$. The last term in eq. \ref{eqNS} is the non-dissipative part of the stress 
tensor \cite{Bray} (for the above dimensionless form see \cite{8}); 
this term represents hydrodynamic interactions. The constant 
$C = \sigma \cdot \xi_{int}/(a \cdot \eta \cdot M)$ depends on the fluid properties, such as the shear 
viscosity, $\eta $, 
interfacial tension, 
$\sigma \approx k\varphi _{eq}^2 / \xi _{int} $, and diffusivity $a M$. The value 
of $C$ determines the importance of hydrodynamic interactions for the specific fluid. 
For a fluid with a high viscosity, where $C<<1$, hydrodynamic interactions can be neglected. 
In this work, we set $C=0$; therefore, the velocity profile in our system is 
determined by the imposed pressure gradient. Thus, advection in a shear flow and diffusion of the 
fluids to the more wettable A or B domains control the evolution of the order parameter in the system 
and are responsible for the observed rich behavior \cite{note}. 

Equations (\ref{eq1})-(\ref{eq3}) are discretized and solved numerically by a cell dynamic system 
method \cite{9} on a 120x40x40 grid. The following boundary conditions on the walls of the 
channel are imposed: $\left. {\frac{\partial \mu }{\partial n}} 
\right|_{wall} = 0$,  $\left. {\frac{\partial 
\varphi }{\partial y}} \right|_{y = o,h} = 0$ and $\left. {\frac{\partial \varphi (\vec {s})}{\partial 
z}} \right|_{z = o,h} = \left. {k^{ - 1}\int {d\vec {s}_i \left[ {V(\vec 
{s}_i )\left( {\varphi (\vec {r}) - \tilde {\varphi }(\vec {s}_i )} \right)} 
\right]} } \right|_{\vec {r} \to \vec {s}} $. The last condition arises explicitly from the minimization 
of free energy in the presence of the substrate potential. At the entry of the channel, 
we have two-stream flow; at the exit, we assume free draining flow, i.e., 
$\left. {\frac{\partial \varphi }{\partial x}} \right|_{x = L} = 0$. For the velocity field, 
we assume no-slip boundary condition on the walls \cite{no-slip}. 

We consider the order parameter evolution in the channel at different values 
of $H$. At low $H$ (low velocities), local thermodynamics dominate, and the fluid just mimics 
the underlying checkerboard pattern, with small distortions in 
$\varphi$ caused by the imposed flow. Higher velocities lead to more 
dramatic changes in $\varphi$ and yield the complex interfaces between the 
A/B fluids shown in Figures 2 and 3. This behavior occurs when the scale of spatial distortions 
in $\varphi$ within the center of the channel (where the Poiseuille flow exhibits the 
maximum velocity, $v^{\max })$ is comparable to the length of the patch, 
i.e., when $v^{\max }t_{diff}^l \approx l$, where $t_{diff}^l \approx l^2 $ is the 
characteristic diffusion time over the patch length $l$. This estimate 
yields a value of $H \approx 10^{ - 4}$. For higher values of H, 
the fluid flows through the middle of the channel with essentially no 
distortion, while near the top and the bottom substrates, $\varphi (r,t)$ is 
governed by the patches' wetting properties (see Fig. 3e).

For the intermediate $H$ values, a competition between the preferential wetting 
interactions and the imposed flow leads to fascinating behavior. On one 
hand, the wetting effects cause the fluids to diffuse to the respective 
patches to minimize the free energy and there is the general tendency to 
minimize interfacial regions between the A and B phases. On the other hand, 
the imposed flow carries fluid away from the favorable patches. If both 
substrates contained just the first half of the checkerboard (a single A and 
B patch), these patches would simply ``switch'' the location of the fluids, 
and the imposed flow would move the switched fluids along the channel. The 
presence of the second set of patches interrupts this flow because both the 
A and B streams again confront incompatible domains. In three dimensions, 
each component can avoid the second unfavorable region by diffusing into the 
bulk. However, recall that the first yellow (B) patch is on the blue (A) 
side (see Fig. 2). Thus, the B fluid can extend only so far into the 
incompatible domain (similar arguments hold for the A fluid). Each fluid 
forms ``arms'' that reach from the top and bottom of the walls; these arms 
can join and pinch off to a form a bubble. Our coarse-grained modeling 
allows such a topology change without any \textit{ad hoc} rules. Figures 2a-c show the order 
parameter distribution at the ``front'' of the microchannel; the same behavior occurs 
for the A fluid at the ``back'' of the channel (Figs. 2d-f).

\begin{figure}[htbp]
\centerline{\includegraphics[width=3.42in,height=1.82in]{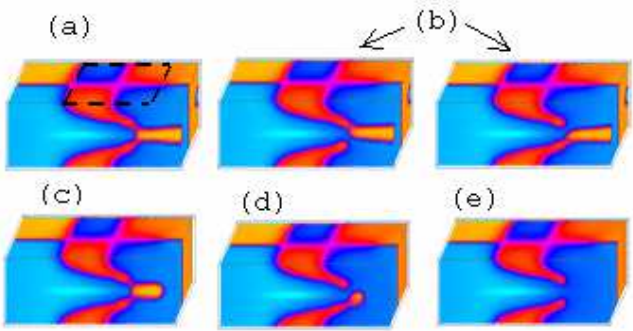}}
\label{fig3}
\caption {Order parameter distribution at steady state for different 
H: a) $H = 2.6 \cdot 10^{ - 4}$, b) $H = 2.7 \cdot 10^{ - 4}$, c) $H = 2.79 
\cdot 10^{ - 4}$, d) $H = 3.2 \cdot 10^{ - 4}$ , e) $H = 3.44 \cdot 10^{ - 
4}$. All other parameters are the same as in Fig 2.}
\end{figure}

Between the limiting cases in Figures 3a and e for relatively low and high 
$H$, respectively, three different types of behavior have been observed: two types of 
periodic behavior and a
time-independent state asymmetric with respect to the top and bottom substrates (Fig. 3b); 
which of the two morphologies shown in Fig. 3b is actually realized for fixed $H$ depends on the noise in 
the system \cite{7}. The periodic cases exhibit ``symmetric'' (Fig. 3c) and ``asymmetric'' 
oscillations (Fig. 3d). The arms in the figures and all the periodic 
behavior develop mainly near the sidewalls. In the middle of the box, the 
Pouseuille velocity field has a maximum and advection prevails. Near the 
wall, however, the velocity is much smaller and diffusion dominates, 
allowing the arms to move upward (downward) and join.

To analyze the complex dynamics, we define a parameter that characterizes 
the integrated changes in $\varphi 
(\mathord{\buildrel{\lower3pt\hbox{$\scriptscriptstyle\rightharpoonup$}}\over 
{r}} ,t)$ near the sidewalls:

\begin{equation}
\label{eq4}
B_i (t) = \frac{1}{V_i }\int {d\vec {r}} \left| {\varphi (\vec {r},t) - 
\varphi (\vec {r},0))} \right|,
\end{equation}

\noindent
here \textit{i = top, bot} indicates whether we integrate over the top (see red dashed box in 
Fig.1) or bottom half of the sidewall region of volume $V_i $.(We choose the 
thickness of this region as $h / 8$.)
\begin{figure}[htbp]
\centerline{\includegraphics[width=3.3in,height=2.5in]{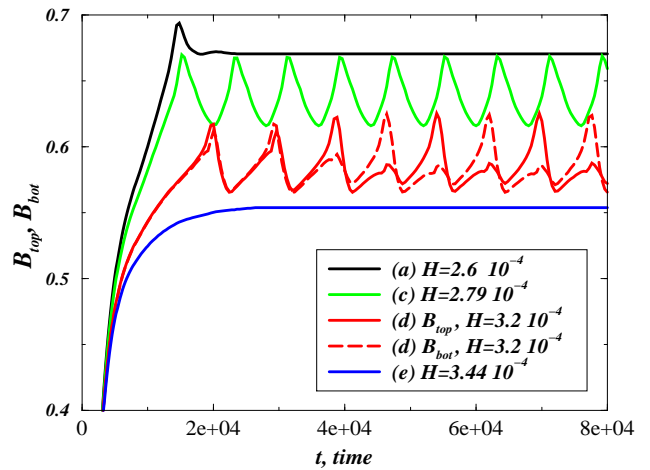}}
\label{fig4}
\caption{Evolution of $B_{top} $ and $B_{bot} $ for cases shown in 
Fig 3, a, c, d and e}
\end{figure}

\noindent

The evolution of $B_{top} (t)$ and $B_{bot} (t)$ for different values of $H$ is 
shown in Fig. 4. Note that $B_{top} (t) = B_{bot} (t)$ for all the symmetric 
cases (Figs. 3 a, c and e). A case of asymmetric oscillations, where $B_{top} (t) \ne B_{bot} (t)$, 
is plotted in red. The 
maxima in the curves correspond to the largest distortions (where the 
bubbles are biggest); the minima correspond to the structure where the arms 
are separated by the greatest distance. The two curves for the asymmetric 
case (Fig. 3d) are similar to each other, but there is a phase shift between 
them. Each of these red curves displays two maxima and two minima in the periodic state. 
At early times, the system's behavior is similar to the 
symmetric case, but at some time, spontaneous symmetry-breaking occurs. The 
length of time before steady-state is reached in the asymmetric case depends 
on degree of noise introduced in the strength of fluid-substrate 
interaction  \cite{7}. The fact that one of the peaks becomes weaker than the 
other indicates that instead of the arms growing equally from both 
substrates and simultaneously forming a bubble in the middle, the top arm, 
for example, grows faster and contributes more to the bubble (higher 
maximum) than the bottom arm (smaller maximum). But for the next bubble, the 
situation is reversed, so the whole period encompasses both maxima; 
this period is roughly twice that of the symmetric case.

We also examined the system response as we change $ H $ dynamically. For each value 
of $H$, the steady-state value of B$_{i}$, or the maximum and minimum in the 
oscillatory regimes, are shown on the bifurcation diagram in Fig. 5. By 
abruptly increasing $H$ from below $H_1 $ to above $H_6 $, the system 
switches from the symmetric time-independent low-velocity regime (as in Fig. 
3a) to the time-independent high-velocity regime (as in Fig. 3e). 
Correspondingly, by abruptly decreasing $H$ from above $H_6 $ to below $H_1$, 
the system switches from the Fig. 3e to the Fig. 3a regime. But the 
transition region between these two points is highly complicated, and 
incorporates complex bifurcations between all the different states. For $H_1 
\le H \le H_6 $, the behavior of the system depends on the starting value of 
$H$, the direction of change (increasing or decreasing) and the magnitude of 
the change. For example, as we gradually increase $H$ from below $H_1 $ (see 
red curve in Fig 5), the time-independent symmetric behavior (as in Fig. 3a) 
becomes unstable and symmetry breaking occurs at the bifurcation point $H_1 $, 
giving rise to the one of the two possible time-independent 
asymmetric types of patterns (as in Fig. 3b). Once the symmetry of the 
system is broken, we find that it is no longer possible to form arms that 
grow symmetrically from the top and bottom substrate and subsequently form 
bubbles that are symmetric with respect to top/bottom. Further gradual 
increases in $H$ lead to the asymmetric oscillations (as in Fig. 3d) for $H 
> H_5 $. For $H_5 < H < H_6 $, we only observe asymmetric oscillations, 
independent of the direction of change in $H$. Inside the parameter region 
$H_2 < H < H_5 $, we can observe all the different types of behavior shown 
in Figs. 3b-d. Gradually decreasing $H$ leads to a transition from 
asymmetric to symmetric oscillations, and then to the time independent 
asymmetric state (blue curve) \cite{10}. Figure 5 clearly shows that the system 
displays hysterisis. 

\begin{figure}[htbp]
\centerline{\includegraphics[width=3.43in,height=2.4in]{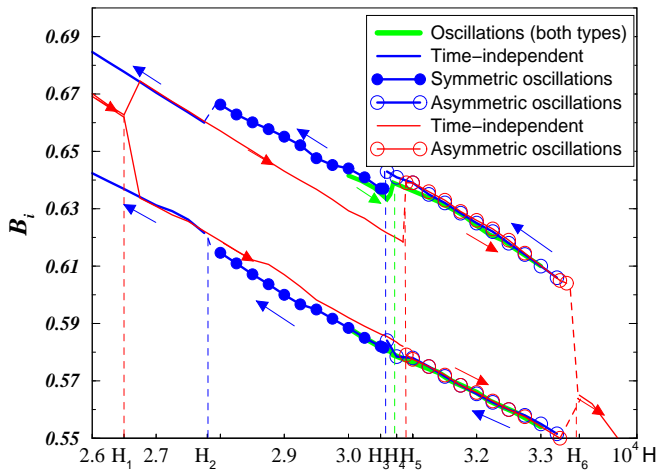}}
\label{fig5}
\caption{Bifurcation diagram. Open circles mark asymmetric 
oscillations, filled circles represent symmetric oscillations, and straight 
lines indicate time-independent behavior (except green lines that depict 
both symmetric ($H < H_4 )$ and asymmetric ($H > H_4 )$ oscillations). All 
other parameters (except $H)$ are the same as in Fig 2. In the oscillatory 
regimes, the top and bottom curves correspond to the maximum and minimum of 
$B_i $.}
\end{figure}

\noindent

The non-decaying, time-periodic behavior in a simple binary fluid driven 
through the microchannel arises from interactions between the fluid and the 
patterned substrate, which introduces non-linearity into the dynamical 
system. These interactions act as an ``activator'' in reaction-diffusion 
systems \cite{2} and are responsible for the positive feedback. We note that the 
periods of the oscillations and the positions of the bifurcation points are 
dependent on the strength of the fluid-substrate interactions and the patch 
length. Thus, the system dynamics can potentially be controlled by varying 
these chemical features. In particular, other choices of patterns can 
potentially lead to new spatiotemporal patterns and dynamical behavior. 

\textbf{Acknowledgements }The authors acknowledge helpful discussions with 
Dr. Sten Ruediger. Supported by the ONR and NSF.


\begin{thebibliography}{99}
\bibitem{7} We introduce noise in the strength of the interaction at each point of the 
checkerboard pattern, i.e.  $V(\vec {s}) = V(1 + 0.03 \zeta (\vec {s})),$ where the 
independently distributed random numbers $\zeta (\vec {s})$ lie in [-1,1).
\bibitem{note} We have, however, carried out 
simulations with hydrodynamic interactions (where we set $C=10$). We observed qualitatively similar periodic 
droplet formation, but the quantitative characteristics (the actual value of the imposed 
pressure gradient when periodic droplet formation occurs, the period and amplitude of oscillations) 
were altered by the actual value of $C$.
\bibitem{10} The period of the oscillations consists of three characteristic times: 
$\tau _1 $, for two retracted arms to reach each other, $\tau _2 $ , for the 
diffusive growth of a droplet, and $\tau _3 $, the time during which the 
droplet moves away from the second patch. The first time, $\tau _1 $, can be 
estimated as the time for advective motion of the arms along a distance 
comparable to the patch length (near the sidewall where velocity is very 
small). Increases in $H$ lead to increases in $\tau _1$ because the distance 
over which the arms are the most separated is larger for higher H. The time 
scales $\tau _2 $ and $\tau _3 $ decrease with $H$. As a function of H, the 
period for symmetric oscillations is almost constant, while for asymmetric 
oscillations, it grows rapidly with increasing H. In the latter case, the 
arm growth from the substrate ($\tau _1 )$ provides the main contribution to 
the total period.
\end{thebibliography}
\end{document}